\documentclass[aip,sd,amsmath,amssymb,reprint]{revtex4-1}

\usepackage{graphicx}
\usepackage{dcolumn}
\usepackage{bm}
\usepackage{epstopdf}
\usepackage{xcolor}
\usepackage{natbib}
\usepackage{subfigure} 
\begin{document}

\title{Spontaneous emergence of non-planar electron orbits\\during direct laser acceleration by a linearly polarized laser pulse}


\author{A. V. Arefiev}

\author{V. N. Khudik}
\affiliation{Institute for Fusion Studies, The University of Texas, Austin, Texas 78712, USA}

\author{A. P. L. Robinson}
\affiliation{Central Laser Facility, STFC Rutherford-Appleton Laboratory, Didcot OX11 0QX, United Kingdom}

\author{G. Shvets}
\affiliation{Institute for Fusion Studies, The University of Texas, Austin, Texas 78712, USA}

\author{L. Willingale}
\affiliation{University of Michigan, 2200 Bonisteel Boulevard, Ann Arbor, Michigan 48109, USA}

\date{\today}

\begin{abstract}
An electron irradiated by a linearly polarized relativistic intensity laser pulse in a cylindrical plasma channel can gain significant energy from the pulse. The laser electric and magnetic fields drive electron oscillations in a plane making it natural to expect the electron trajectory to be flat. We show that strong modulations of the relativistic $\gamma$-factor associated with the energy enhancement cause the free oscillations perpendicular to the plane of the driven motion to become unstable. As a consequence, out of plane displacements grow to become comparable to the amplitude of the driven oscillations and the electron trajectory becomes essentially three-dimensional, even if at an early stage of the acceleration it was flat. The development of the instability profoundly affects the x-ray emission, causing considerable divergence of the radiation perpendicular to the plane of the driven oscillations, while also reducing the overall emitted energy.
\end{abstract}

\maketitle

\section{Introduction}

Generation of energetic electrons is a key feature of ultra-intense laser-plasma interactions and it has been successfully employed in a variety of applications, including radiation~\cite{Park2006,Kneip2008} and particle sources~\cite{Pomerantz2014, Schollmeier2015, Chen2015}. The laser-plasma interaction and thus the mechanism responsible for electron acceleration strongly depend on the duration of the laser pulse and the plasma density~\cite{Wilks1992, Pukhov1999, Esarey2009, Kemp2012}. In the case of an underdense plasma, the laser pulse can propagate through the plasma and its propagation is typically accompanied by cavitation of the electron density~\cite{Gahn1999, Mangles2005, Sentoku2006, Willingale2013}. 

In this paper, we focus on the regime where the laser pulse is sufficiently long to establish a slowly evolving channel~\cite{Willingale2011}. This implies that the duration of the laser pulse significantly exceeds the period of plasma oscillations. This regime can naturally arise due to target ablation during the main pulse or during the pre-pulse even if the initial target density is over-critical. The channel produced by expelling some of the electrons generates a transverse quasi-static electric field. It has been shown that the presence of such a field can facilitate the electron energy gain directly from the laser, making it possible to generate copious electrons with energies exceeding the ponderomotive potential~\cite{Gahn1999, Pukhov1999, Mangles2005, Arefiev2012}. The energy enhancement in combination with the strong acceleration experienced by the electron in the channel can also be beneficial for x-ray emission, as has been demonstrated in experiments with gas jets~\cite{Kneip2008}. 

Both the relevance of the discussed regime to a range of applications and its potential to generate energetic electrons and to boost the x-ray emission have renewed interest in exploring mechanisms of electron acceleration in plasma channels~\cite{Arefiev2012, Liu2013, Robinson2013, Krygier2014, Liu2015}. When analyzing the dynamics of electrons accelerated by a linearly polarized laser pulse inside a channel, i.e. the so-called direct laser acceleration regime, it might appear reasonable to treat their trajectories as flat. Indeed, the only oscillating electric field transverse to the axis of the channel is the laser electric field. For an electron starting its motion exactly on the axis of the channel, the laser electric field would drive strong transverse oscillations, while the Lorentz force would cause longitudinal motion, thus producing a flat electron trajectory.

In this paper, we show that such flat electron trajectories can be inherently unstable with respect to small transverse displacements perpendicular to the plane of the driven motion. The cause of the instability is the coupling through the relativistic $\gamma$-factor of the driven and free transverse oscillations in a cylindrical channel. The oscillations driven by the laser induce strong modulations of the relativistic $\gamma$-factor. As a result, the frequency of the free oscillations is also modulated, which makes it possible for these oscillations to become parametrically unstable under appropriate conditions. We show that such conditions are met after the relativistic $\gamma$-factor or the electron energy associated with the driven motion becomes significantly enhanced. The energy enhancement that takes place in an ion channel due to the presence of the static electric field~\cite{Pukhov1999} has a well-pronounced threshold determined by the wave amplitude and the ion density in the channel~\cite{Arefiev2012, Arefiev2014}. We have determined that even small off-axis displacements perpendicular to the plane of the driven motion quickly grow above this threshold. Therefore, considerable electron energy enhancement in a cylindrical channel necessarily leads to a fully three-dimensional electron trajectory.  

Presented here analysis bridges the gap in understanding of electron dynamics in cylindrical channels. Previously, the parametric instability and its impact on electron acceleration has been studied using only slab-like two-dimensional ion channels~\cite{Arefiev2012, Arefiev2014} . The main advantage of the two-dimensional setup is that it allows to spatially decouple the oscillations driven by the laser and the oscillations caused by the electrostatic field of the channel. This aspect makes the problem analytically tractable and thus greatly facilitates the analysis of the electron acceleration mechanism. On the other hand, the results obtained for the slab-like channels cannot be directly applied to the more relevant case of cylindrical channels. In a cylindrical channel, the influence of the laser electric field and the field of the channel cannot, in principle, be decoupled. Our analysis shows that non-planar trajectories arise due to this coupling, with the electron energy enhancement serving as the trigger.

We also apply the newly gained insight into the electron dynamics to analyze the x-ray emission by the accelerated electrons. The development of the three-dimensional motion in the channel causes a considerable divergence of the radiation perpendicular to the plane of the driven oscillations, while, at the same time, it reduces the overall emitted energy. Therefore, the three-dimensional aspect of the electron motion must be taken into account when analyzing the x-ray generation during the direct laser acceleration. 

\section{Basic model} \label{Basic_Model}

\begin{figure}
	\centering
	\includegraphics[width=0.9\columnwidth]{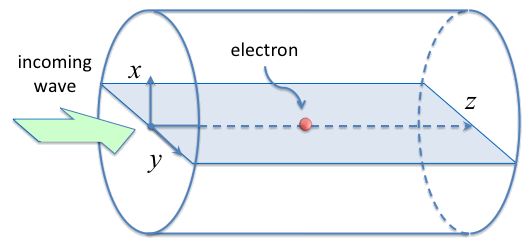} 
  \caption{Cylindrical setup: a single electron is irradiated by a plane electromagnetic wave in a cylindrical ion channel. The laser electric field is polarized in the plotted $(y,z)$-plane.} \label{Figure0}
\end{figure}

In order to examine the effect of the parametric instability, we consider just a single electron irradiated by a plane electromagnetic wave in a fully evacuated cylindrical ion channel, as shown in Fig.~\ref{Figure0}. An important difference in this setup from those used in Refs.~[\onlinecite{Arefiev2012}] and [\onlinecite{Arefiev2014}] is the configuration of the static electric field. The advantage of the single electron model is that the fields acting on the electron can be treated as given.


The incoming wave irradiating the electron is a plane linearly polarized electromagnetic wave propagating along the channel with a phase velocity $v_{ph}$ equal to the speed of light. The main effect investigated in this work is not very sensitive to the value of $v_{ph}$, so we set $v_{ph} = c$ only to simplify the analysis. It is convenient to use a Cartesian system of coordinates $(x,y,z)$, with the $z$-axis directed along the axis of the channel and in the direction of the wave propagation. Without any loss of generality, we set the wave electric and magnetic fields to be directed along the $y$ and $x$-axis, respectively. To describe the wave propagation, we introduce a normalized vector potential that has only a $y$-component. Its amplitude $a$ is only a function of a single dimensionless phase variable
\begin{equation}
\xi = \omega \left( t - z/c \right),
\end{equation}
where $\omega$ is the frequency of the wave and $c$ is the speed of light. The electric and magnetic fields of the wave are then given by
\begin{eqnarray} \label{laser-field}
&& E_y = - \frac{m_e \omega c}{|e|} \frac{d a}{d \xi}, \\
&& B_x = -E_y,  \label{laser-field-2}
\end{eqnarray}
where $e$ and $m_e$ are the electron charge and mass. 

The electric field of the channel has both $x$ and $y$ components. Assuming for simplicity that the channel is a uniform positively charged cylinder that consists of immobile singly-charged ions of density $n_0$, we find that 
\begin{eqnarray}
&& E_x = 2 \pi n_0 |e| x = \left. m_e \omega_{p0}^2 x \right/ 2 |e|, \label{E-chan-x}\\
&& E_y = 2 \pi n_0 |e| y = \left. m_e \omega_{p0}^2 y \right/ 2 |e|, \label{E-chan-y}
\end{eqnarray}
where $\omega_{p0} = \sqrt{4 \pi n_0 e^2/m_e}$ is the plasma frequency. 

The electron dynamics obeys the following general equations of motion:
\begin{eqnarray}
&& \frac{d {\bf{p}}}{dt} = - |e| {\bf{E}} - \frac{|e|}{\gamma m_e c} \left[ {\bf{p}} \times {\bf{B}} \right], \label{eq-mot-1}\\
&& \frac{d {\bf{r}}}{dt} = \frac{{\bf{p}}}{\gamma m_e}, \label{eq-mot-2}
\end{eqnarray}
where ${\bf{p}}$ is the electron momentum, ${\bf{r}}$ is the electron displacement from the axis of the channel, and 
\begin{equation}
\gamma = \sqrt{1 + {\bf{p}}^2/m_ec^2}
\end{equation}
is the relativistic factor. The fields ${\bf{E}}$ and ${\bf{B}}$ are superpositions of the fields of the wave [Eqs.~(\ref{laser-field}) and (\ref{laser-field-2})] and the fields of the channel [Eqs.~(\ref{E-chan-x}) and (\ref{E-chan-y})].

It is important to point out that the amplitude of the transverse oscillations has a hard upper limit regardless of the wave profile. One can verify that Eqs.~(\ref{eq-mot-1}) and (\ref{eq-mot-2}) have an integral of motion,
\begin{eqnarray}
I = \gamma - \frac{p_z}{m_e c} + \frac{\omega_{p0}^2}{4 c^2} \left( x^2 + y^2 \right) = \mbox{const}, \label{R_def}
\end{eqnarray}
that relates the amplitude of the oscillations across the channel and the electron $\gamma$-factor. The value of $I$ is determined by the initial conditions, so that we have $I = 1$ for an on-axis electron that is initially at rest. We find from Eq.~(\ref{R_def}) that the amplitude of the electron oscillations cannot exceed 
\begin{eqnarray}
r_{\max} = \frac{\lambda}{\pi} \frac{\omega}{\omega_{p0}} \label{R_max_2}
\end{eqnarray}
for $I = 1$. The amplitude of the oscillations approaches $r_{\max}$ as $p_z/m_e c \rightarrow \infty$ and $\gamma - p_z/m_e c \rightarrow 0$. It follows from Eq.~(\ref{R_def}) that the constant $I$ is also close to unity for an initially displaced electron if the displacement is much less than $2c/\omega_{p0}$. Therefore, $r_{\max}$ serves as an upper limit for the transverse oscillations in this case as well.

In what follows, we consider an incoming wave whose amplitude monotonically increases from zero to its maximum value of $a_0$ over many wave periods and then remains constant. The specific shape used to numerically solve the equations of motion is
\begin{equation}
a(\xi) =
   \begin{cases}
      a_0 \exp[-(\xi-\xi_0)^2/2 \sigma^2] \sin(\xi),  & \text{for $\xi <  \xi_0$;}\\
      a_0 \sin(\xi),  & \text{for $\xi \geq  \xi_0$}
   \end{cases} \label{a}
\end{equation}
where the parameters $\xi_0 = 50$ and $\sigma = 20$ determine the ramp-up of the amplitude.

Our objective is to determine the key features of electron dynamics in the channel and the sensitivity of electron trajectories to displacements off the axis of the channel. The electric and magnetic fields of the wave with the chosen polarization drive electron oscillations in the $(y,z)$-plane. Therefore, the displacements of particular interest are the displacements along the $x$-axis and out of the plane of the driven oscillations. In what follows, we consider an electron that is initially at rest and we distinguish two cases of interest. In the first case, the electron is located exactly on the axis of the channel prior to the arrival of the wave, so that $x=y=z = 0$ at $t=0$. In the second case, the electron is initially displaced along the $x$-axis, so that $x= \Delta x$ and $y=z = 0 $ at $t=0$.

\section{Driven oscillations in the (y,z)-plane} \label{Section-Driven}

\begin{figure}
  \includegraphics[width=0.9\columnwidth]{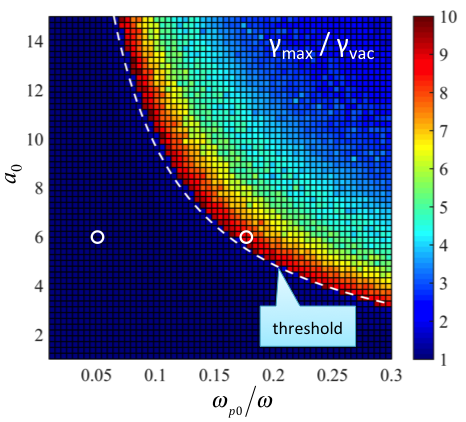}
  \caption{Maximum $\gamma$-factor ($\gamma_{\max}$) achieved by an electron irradiated by a plane electromagnetic wave with maximum amplitude $a_0$ in a cylindrical channel with ion density $n_0/n_{\mbox{crit}} = \omega_{p0}^2/\omega^2$. The relativistic factor $\gamma$ is normalized to $\gamma_{\mbox{vac}} =  1 + a_0^2/2$. Initially, the electron is at rest on the axis of the channel.} \label{Figure_scan}
\end{figure}

In this section, we consider the first case of interest outlined in the previous section: the electron is initially at rest and it is located on the axis of the channel. This electron experiences no force directed along the $x$-axis and, therefore, its trajectory will remain flat as it performs driven oscillations in the $(y,z)$-plane shown in Fig.~\ref{Figure0}.

\begin{figure}
	\centering
    	\subfigure{\includegraphics[width=0.9\columnwidth]{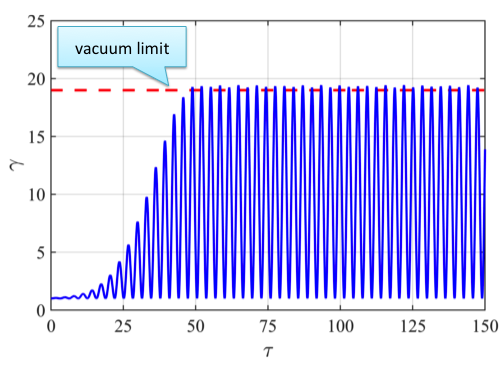} } \\
	\subfigure{\includegraphics[width=0.9\columnwidth]{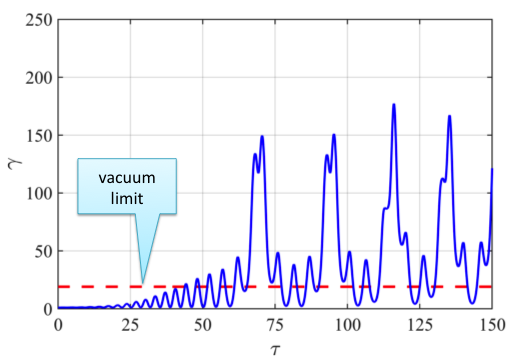}} 
  \caption{Electron $\gamma$-factor for $a_0 = 6$ and ion densities below (upper panel) and above (lower panel) the threshold shown in Fig.~\ref{Figure_scan}. The densities correspond to $\omega_{p0} / \omega = 0.05$ (upper panel) and $\omega_{p0} / \omega = 0.163$ (lower panel). The parameters corresponding to the upper and lower panels are marked with empty circles in Fig.~\ref{Figure_scan}. The $\gamma$-factor is given as a function of the proper time defined by Eq.~(\ref{t-tau}). The dashed line marks $\gamma_{\mbox{vac}}$, which is the maximum value of $\gamma$ in the absence of the channel fields.} \label{Figure1_1}
\end{figure}

We substitute the expressions for the fields of the laser [Eqs.~(\ref{laser-field}) and (\ref{laser-field-2})] and the fields of the channel  [Eqs.~(\ref{E-chan-x}) and (\ref{E-chan-y})] into Eqs. (\ref{eq-mot-1}) and (\ref{eq-mot-2}) to obtain the following  set of equations for the electron motion in the $(y,z)$-plane:
\begin{eqnarray}
&& \frac{d p_y}{dt} = - \frac{1}{2} m_e \omega_{p0}^2 y + m_e \omega c \left( 1 - \frac{p_z}{\gamma m_e c} \right) \frac{d a}{d \xi}, \label{main-2}\\
&& \frac{d p_z}{dt} = m_e \omega c \frac{p_y}{\gamma m_e c} \frac{d a}{d \xi}, \label{main-3}\\
&& \frac{d y}{dt} = \frac{p_y}{\gamma m_e}, \label{main-5}\\
&& \frac{d \xi}{dt} = \omega \left( 1 - \frac{p_z}{\gamma m_e c} \right). \label{main-6}
\end{eqnarray}
Note that the last equation is equivalent to the equation $dz/dt = p_z/ \gamma m_e$. We use the equation for $\xi$ instead, since the vector potential $a$ can be expressed as just a function of $\xi$. Driven electron oscillations across the channel were first considered in Ref.~[\onlinecite{Pukhov1999}] and then later investigated in more detail in Refs.~[\onlinecite{Arefiev2012}] and [\onlinecite{Arefiev2014}]. In this section, we follow the analysis of Ref.~[\onlinecite{Arefiev2014}].

The electron motion in the channel is influenced by two parameters: $a_0$, representing the wave amplitude, and $\omega_{p0}/\omega$, representing the ion density in the channel. We have performed a parameter scan, solving Eqs.~(\ref{main-2}) - (\ref{main-6}) numerically for a range of values of $a_0$ and $\omega_{p0}/\omega$ and the wave specified by Eq.~(\ref{a}). The maximum $\gamma$-factor, denoted by $\gamma_{\max}$, that the electron achieves moving through the channel for different sets of parameters $a_0$ and $\omega_{p0}/\omega$ is shown in Fig.~\ref{Figure_scan}. On the other hand, the maximum $\gamma$-factor that the electron can reach moving in the wave without the additional influence of the field generated by the ions is
\begin{equation}
\gamma_{\mbox{vac}} = 1 + a_0^2/2.
\end{equation}
Fig.~\ref{Figure_scan} shows the enhancement of $\gamma_{\max}$ relative to $\gamma_{\mbox{vac}}$.

As evident from this figure, there is a well pronounced energy enhancement threshold. Below the threshold, the maximum electron $\gamma$-factor is comparable to $\gamma_{\mbox{vac}}$, whereas, above the threshold, the maximum $\gamma$-factor  significantly exceeds $\gamma_{\mbox{vac}}$. The threshold is determined by a single dimensionless combination $a_0 \omega_{p0} / \omega$~[\onlinecite{Arefiev2014}]. The dashed curve in Fig.~\ref{Figure_scan} that follows the threshold corresponds to $a_0 \omega_{p0} / \omega = 0.972$. It must be noted that the exact location of the threshold depends on the initial displacement and momentum of the electron~[\onlinecite{Arefiev2014}] and can also be influenced by the ramp-up of the wave amplitude. 

Figure~\ref{Figure1_1} shows profiles of the electron $\gamma$-factor below and above the threshold for parameters marked with open circles in Fig.~\ref{Figure_scan}. The $\gamma$-factor is given as a function of the dimensionless proper time $\tau$ defined as
\begin{equation} \label{t-tau}
	\frac{d \tau}{d t} = \frac{\omega}{ \gamma}.
\end{equation}
This representation is convenient for the analysis that follows in the next section. In both cases, the wave amplitude is the same, $a_0 = 6$, so the threshold is encountered by increasing the ion density. The upper panel in Fig.~\ref{Figure1_1} corresponds to $\omega_{p0} / \omega = 0.05$ (left circle in Fig.~\ref{Figure_scan}). This value is well below the threshold for $a_0 = 6$ and, therefore, $\gamma_{\max}$ is roughly the same as $\gamma_{\mbox{vac}}$. Moreover, $\gamma(\tau)$ has a well defined period. The lower panel in Fig.~\ref{Figure1_1} corresponds to $\omega_{p0} / \omega = 0.163$ (right circle in Fig.~\ref{Figure_scan}) that is slightly above the threshold. Not only there is a dramatic increase of $\gamma_{\max}$, but there is also a significant change in the profile of $\gamma(\tau)$. The $\gamma(\tau)$ now has sharp semi-periodic spikes in addition to the frequent oscillations that are also seen below the threshold (upper panel of Fig.~\ref{Figure1_1}). Even though the profiles of $\gamma(\tau)$ in Fig.~\ref{Figure1_1} are only a segment of the time interval ($0 \leq \tau \leq 600$) used to generate the parameter scan shown in Fig.~\ref{Figure_scan}, they capture the key changes in the evolution of the the relativistic $\gamma$-factor.

\section{Free oscillations along the x-axis} \label{Section-Free}

In this section, we consider the second case of interest, with the electron being initially slightly displaced along the $x$-axis. The displacement initiates electron motion in the $x$-direction, since the restoring force along the $x$-axis no longer vanishes.

To take into account the motion along the $x$-axis, Eqs.~(\ref{main-2}) - (\ref{main-6}) should simply be supplemented by the following two equations that are $x$-components of Eqs.~(\ref{eq-mot-1}) and (\ref{eq-mot-2}):
\begin{eqnarray}
&& \frac{d p_x}{dt} = - \frac{1}{2} m_e \omega_{p0}^2 x, \label{main-1}\\
&& \frac{d x}{dt} = \frac{p_x}{\gamma m_e}. \label{main-4}
\end{eqnarray}
We combine Eqs.~(\ref{main-1}) and (\ref{main-4}) into a single equation, 
\begin{eqnarray}  \label{main-4-2}
 \frac{d^2 x}{d \tau^2}+ \Omega^2 x = 0,
\end{eqnarray}
that resembles that of an oscillator with a natural frequency
\begin{equation}
\Omega \equiv \sqrt{\frac{\gamma}{2}} \frac{\omega_{p0}}{\omega}.
\end{equation}
Here $\tau$ is the the dimensionless proper time defined by Eq.~(\ref{t-tau}).

Electron oscillations described by Eq.~(\ref{main-4-2}) and the driven motion in the $(y,z)$-plane are coupled only via the relativistic $\gamma$-factor. This means that the driven motion described in Sec.~\ref{Section-Driven} remains essentially unaltered for as long as the amplitude of the oscillations, and the momentum associated with them, remain small. In order to find out whether the oscillations can grow, we analyze Eq.~(\ref{main-4-2}) assuming that the $\gamma$-factor is determined only by the driven motion.

The amplitude of the oscillations can only grow due to modulations of the natural frequency $\Omega$. In an axially uniform channel that we are considering, $\Omega$ is modulated by the $\gamma$-factor. The modulations of $\gamma$ induced by the driven electron motion in the $(y,z)$-plane are shown in Fig.~\ref{Figure1_1} for ion densities below and above the energy enhancement threshold. Below the threshold, $\gamma(\tau)$ consists of periodic oscillations between $\gamma \sim 1$ and $\gamma_{\max}$. We will refer to one such oscillation in this regime as a single modulation. Above the threshold, the profile of $\gamma(\tau)$ is rather different, but one can still distinguish a semi-periodic pattern: there are tall spikes with a comparable delay between them. It is then appropriate in this regime to refer to one such spike as a single modulation.

\begin{figure}
	\centering
    	\includegraphics[width=0.9\columnwidth]{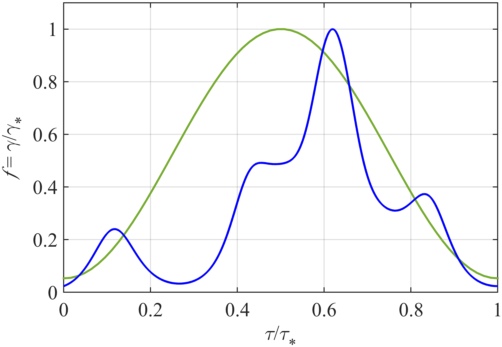} 
  \caption{Normalized modulations of the relativistic $\gamma$-factor at ion densities above (blue curve) and below (green curve) the energy enhancement threshold. Here $\tau_*$ is the period and $\gamma_*$ is the amplitude of the modulations. Their values are different in each case (see Sec.~\ref{Section-Free} for more details).} \label{FigureX}
\end{figure}

To gain insight into electron dynamics caused by the modulations of $\gamma$, we take just a single modulation and assume that it is periodically repeated.  We use $\tau_*$ and $\gamma_*$ to denote the period of the modulations and their amplitude. The $\gamma(\tau)$ from the upper panel of Fig.~\ref{Figure1_1} is represented by one of the peaks shown in Fig.~\ref{FigureX} by a green curve, with $\gamma_* \approx 19$ and $\tau_* \approx 3.15$. The $\gamma(\tau)$ from the lower panel of Fig.~\ref{Figure1_1} is represented by the third spike shown in Fig.~\ref{FigureX} by a blue curve, with  $\gamma_* \approx 177$ and $\tau_* \approx 19.2$.

Now that we have approximated the relativistic $\gamma$-factor by a periodic function, 
\begin{equation}
\gamma(\tau) = \gamma_* f(\tau/\tau_*),
\end{equation}
Eq.~(\ref{main-4-2}) can be rewritten as
\begin{eqnarray}  \label{model_1}
 \frac{d^2 x}{d s^2}+ \Omega_0^2 f(s) x = 0,
\end{eqnarray}
where 
\begin{eqnarray} 
&& s \equiv \tau / \tau_*, \\
&& \Omega_0 \equiv \sqrt{\frac{\gamma_*}{2}} \frac{\omega_{p0}}{\omega} \tau_*, \label{model_2}
\end{eqnarray}
and $f$ is a normalized periodic function that represents the structure of a specific modulation that is chosen to approximate $\gamma$. By definition, the period of $f$ is equal to unity and $0 < f \leq 1$. General properties of Eq. (\ref{model_1}) are well-known and appears in many physics problems. The feature that is of the key importance here is that Eq. (\ref{model_1}) can have exponentially growing solutions for some bands of frequencies $\Omega_0$.

\begin{figure}
	\centering
	\includegraphics[width=0.9\columnwidth]{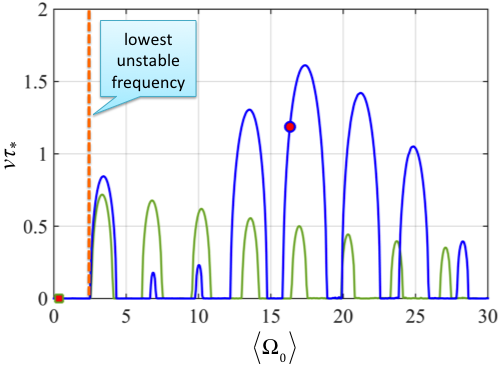}
  \caption{Exponential growth rate $\nu$ of the free oscillations along the $x$-axis for the two modulations shown in Fig.~\ref{FigureX}.  Here $\tau_*$ is the period of the modulations and $\left< \Omega_0 \right>$ is the average frequency of the oscillations defined by Eq.~(\ref{mu}).} \label{FigureX2}
\end{figure}

Treating $\Omega_0$ as a free parameter, we have numerically calculated the exponential growth rate $\nu$ for both modulations from Fig.~\ref{FigureX} over a broad range of $\Omega_0$ values. The corresponding curves are shown in Fig.~\ref{FigureX2} versus the average frequency,
\begin{equation} \label{mu}
\left< \Omega_0  \right>  \equiv \left[ \frac{1}{\tau_*} \int_0^{\tau_*} \Omega_0 f(\tau / \tau_*) d \tau \right]^{1/2},
\end{equation}
to emphasize their similarity. In both cases, there are alternating stable and unstable frequency bands. In the unstable bands, Eq.~(\ref{model_1}) has a solution whose amplitude increases exponentially as $\exp(\nu \tau)$ with every modulation. The lowest unstable band for both curves starts roughly at $\left< \Omega_0 \right> \approx 2.5$.

The apparent similarity of the growth rate curves indicates that the stability of the free electron oscillations along the $x$-axis is not very sensitive to the exact shape of the normalized modulation. The main difference between the modulations of the relativistic $\gamma$-factor below and above the energy enhancement threshold are the respective values of the average frequency $\left< \Omega_0 \right>$. Indeed, taking into account that the modulations shown in Fig.~\ref{FigureX} correspond to $\omega_{p0}/\omega = 0.05$ and $\omega_{p0}/\omega = 0.163$, we find that the average frequency above the threshold, $\left< \Omega_0 \right> \approx 16.3$, significantly exceeds the average frequency below the threshold, $\left< \Omega_0 \right> \approx 0.35$. The average frequency for the modulations below the threshold (marked with a square) is well below the lowest unstable band in Fig.~\ref{FigureX2}. Therefore, electron oscillations along the $x$-axis are stable in this regime and small initial displacements do not grow with time. In contrast with that, the average frequency for the modulations above the threshold (marked with a circle) is in the fifth unstable band. The corresponding exponential growth rate of the oscillations is $\nu \approx 1.19 / \tau_*$.

The key conclusion from the presented analysis is that the change in the modulations of the $\gamma$-factor that accompanies the significant energy enhancement shown in Fig.~\ref{Figure1_1} makes free electron oscillations along the $x$-axis unstable. The main changes that occur are the significant increase of the period and amplitude of the modulations. 

It is important to emphasize that we have assumed strict periodicity when analyzing the modulations shown in Fig.~\ref{Figure1_1}. This assumption holds well if there is no significant enhancement of the $\gamma$-factor. However, as evident from Fig.~\ref{Figure1_1}, the period and amplitude of the modulations are gradually changing in the regime where the $\gamma$-factor is enhanced. Therefore, the unstable frequency bands and the average frequency $\left< \Omega_0 \right>$, shown with a circle in Fig.~\ref{FigureX2}, are not fixed in this case. On one hand, this means that if initially $\left< \Omega_0 \right>$ happens to be between the unstable bands, the subsequent change in the modulations will likely make the oscillations unstable. On the other hand, this raises the question of whether the lack of exact periodicity would stem the growth of the transverse oscillations. We address this in the next section by carrying out fully self-consistent calculations of electron dynamics in the cylindrical channel.

\section{3D electron motion} \label{Section-3D}

\begin{figure}
	\centering
    	\subfigure{\includegraphics[width=0.9\columnwidth]{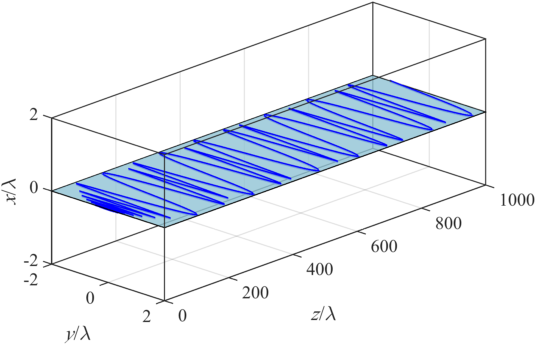} } \\
	\subfigure{\includegraphics[width=0.9\columnwidth]{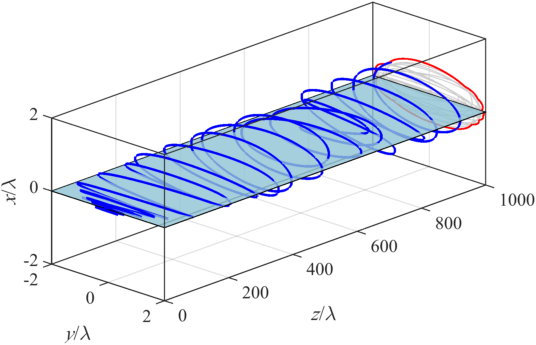}} 
  \caption{Comparison of the electron trajectories with and without the initial displacement (lower and upper panels). The wave electric field is polarized in the $(y,z)$-plane (shown in light-blue). The electron in the lower panel is initially displaced by $\Delta x = 0.1 \lambda$ off the axis of the channel. In both cases, we set $a_0 = 6$ and $\omega_{p0}/\omega = 0.163$.} \label{Figure_example_1}
\end{figure}

In this section, we demonstrate that significant oscillations perpendicular to the plane of the driven oscillations can indeed develop during electron acceleration in a cylindrical channel. To find the three-dimensional electron trajectories, we numerically solve the full system of equations, Eqs.~(\ref{eq-mot-1}) and (\ref{eq-mot-2}), with the wave ramp-up specified by Eq.~(\ref{a}). 

For our first example we take the same channel density (right circle in Fig.~\ref{Figure_scan}) that was used to generate the $\gamma$-factor modulations shown in the lower panel of Fig.~\ref{Figure1_1}. Specifically, the ion density is set at $n_0 / n_{\mbox{crit}} \approx 2.66 \times 10^{-2}$, which corresponds to $\omega_{p0}/\omega = 0.163$. We now compare our two cases: an electron starting its motion on the axis of the channel and an electron that is initially displaced by $\Delta x = 0.1 \lambda$. In both cases, the electron is initially at rest and the wave amplitude is $a_0 = 6$.

The electron trajectories calculated for the two cases are shown in Fig.~\ref{Figure_example_1}. The trajectory of the electron that starts its motion on the axis (upper panel) remains flat throughout the acceleration process. Its motion consists only of the driven oscillations in the $(y,z)$-plane. This is exactly the case that was considered in Sec.~\ref{Section-Driven}. The time evolution of the $\gamma$-factor is shown in Fig.~\ref{Figure1_1}. In this case, the $\gamma$-factor is significantly enhanced and there are strong modulations associated with the enhancement. In Sec.~\ref{Section-Free}, we showed that these modulations can make flat trajectories unstable with respect to small displacements out of the trajectory plane. 

The trajectory of the electron with a small initially displacement (lower panel in Fig.~\ref{Figure_example_1}) confirms the instability, as the electron develops appreciable oscillations along the $x$-axis. Early in the acceleration process, the trajectories with and without the displacement appear visually identical. However, as the electron continues its motion along the channel, the modulations of the $\gamma$-factor induced by the driven oscillations in the $(y,z)$-plane cause the small oscillations along the $x$-axis to grow. A projection of the electron trajectory onto the cross-section of the channel, i.e. onto the $(x,y)$-plane, is also shown in the lower panel of Fig.~\ref{Figure_example_1}. Evidently, the amplitude of the free oscillations reaches roughly half of the amplitude of the driven oscillations. This is a clear manifestation of the instability triggered by the energy enhancement.

\begin{figure}
  \includegraphics[width=0.9\columnwidth]{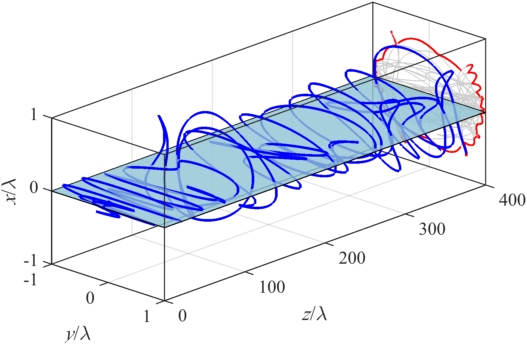}
  \caption{Electron trajectory in a channel with $\omega_{p0}/\omega = 0.326$. The wave amplitude is $a_0 = 6$ and the initial displacement across the channel is $\Delta x = 0.1 \lambda$.} \label{Figure2}
\end{figure}

The development of the oscillations along the $x$-axis that makes the trajectory three-dimensional depends on the initial displacement, ramp-up of the wave amplitude, and the ion density in the channel. Figure~\ref{Figure2} illustrates an electron trajectory in a channel whose ion density is four times higher than that used in the previous example, $n_0 / n_{\mbox{crit}} \approx 1.06 \times 10^{-1}$ ($\omega_{p0}/\omega = 0.326$). All the other parameters, including the initial electron displacement, are exactly the same as those used to generate the trajectory in the lower panel of Fig.~\ref{Figure_example_1}. The electron trajectory shown in Fig.~\ref{Figure2} develops appreciable oscillations along the $x$-axis a lot sooner than in Fig.~\ref{Figure_example_1} and their amplitude becomes essentially equal to the amplitude of the driven oscillations along the $y$-axis.

\begin{figure}
	\centering
    	\subfigure{\includegraphics[width=0.9\columnwidth]{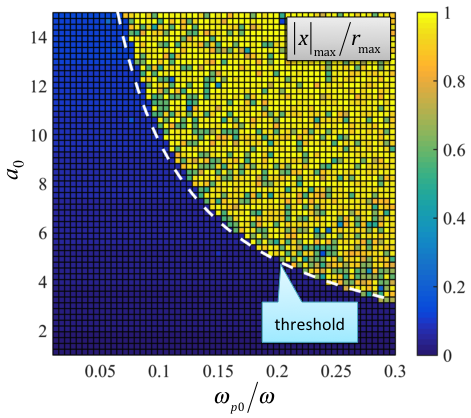} } \\
	\subfigure{\includegraphics[width=0.9\columnwidth]{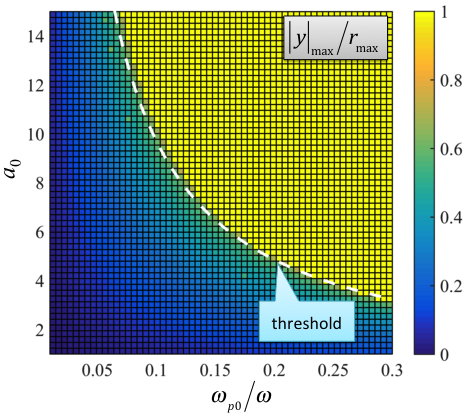}} 
  \caption{Maximum amplitudes of the free (upper panel) and driven (lower panel) oscillations across the cylindrical channel. The amplitudes are normalized to $r_{\max}$ given by Eq.~(\ref{R_max_2}). The initial transverse displacement is set at $\Delta x = 2.7 \times 10^{-3} a_0 \omega/\omega_{p0}$ in order to be much smaller than $r_{\max}$ and $|y|_{\max}$ for all sets of parameters. The dashed curve in both panels marks the energy enhancement threshold shown in Fig.~\ref{Figure_scan}.} \label{Figure_8}
\end{figure}

To determine how the maximum amplitude of the electron oscillations across the channel depends on the wave amplitude and the ion density, we have performed a parameter scan for an initially displaced electron. The results of the scan are given in Fig.~\ref{Figure_8}, where we separately show maximum amplitudes of the free and driven oscillations ($|x|_{\max}$ and $|y|_{\max}$, respectively). The amplitudes are normalized to the maximum amplitude of the transverse oscillations $r_{\max}$ derived in Sec.~\ref{Basic_Model} and given by Eq.~(\ref{R_max_2}). 

Unsurprisingly, there is an enhancement of the driven oscillations along the $y$-axis above the energy enhancement threshold shown with a dashed curve in Fig.~\ref{Figure_8}. The increase of $|y|_{\max}$ manifests the energy increase of the driven oscillations, so the scan results shown in the lower panel of Fig.~\ref{Figure_8} would look identical if we were to consider an electron without any initial displacement instead. 

In stark contrast to this, an initial displacement out of the plane of the driven motion dramatically changes the free electron motion along the $x$-axis. The upper panel of Fig.~\ref{Figure_8} shows that there is a well-pronounced threshold for the enhancement of the free oscillations. This threshold matches the energy enhancement threshold (dashed line), because, as discussed in Sec.~\ref{Section-Free}, the energy enhancement triggers the growth of the electron oscillations along the $x$-axis. The amplitude of the free oscillations grows until it becomes comparable to $r_{\max}$. It is important to point out that the energy enhancement and the increase of the driven oscillations necessarily take place first and only then the free oscillations can be amplified. This explains the checkered patter above the threshold. The pattern becomes more uniform with increased observation time. 

We conclude that, in general, the electron energy enhancement is accompanied by a considerable enhancement of the transverse oscillations across the channel, making the electron trajectory considerably three-dimensional with the maximum displacement given by $r_{\max}$.


\section{Summary and Discussion} \label{Summary}

The analysis presented in this paper bridges a gap in the existing understanding of electron dynamics in cylindrical channels. We have shown that the threshold for the energy enhancement matches the threshold for the onset of the parametric instability in the direction perpendicular to the plane of the driven electron oscillation. As a result, the trajectories of electrons with enhanced energies necessarily become non-planar.

It is worth noting that the energy enhancement takes place first and the development of the parametric instability occurs afterwards. Therefore, limiting the acceleration distance or the length of the channel could be a way to keep electron trajectories flat if the three-dimensional motion is undesirable. The onset of the instability can be further delayed by increasing the wave amplitude $a_0$. Even if the amplification of a small transverse displacement takes the same number of laser oscillations, the corresponding axial distance travelled by the electron increases as $a_0^2$. These considerations might be important to the proposed schemes of the direct laser acceleration of electrons from low density targets by the next generation of high intensity lasers~\cite{Singh2015}. 

\begin{figure}
	\centering
    	\subfigure{\includegraphics[width=0.9\columnwidth]{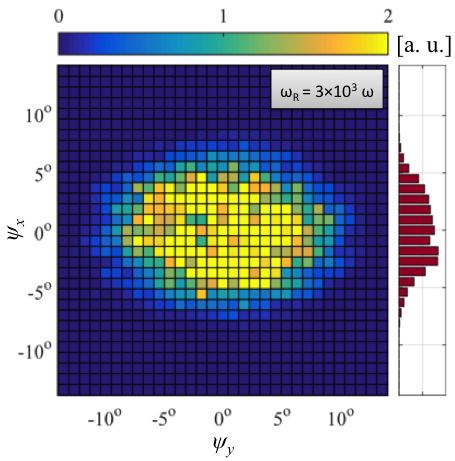} } \\
	\subfigure{\includegraphics[width=0.9\columnwidth]{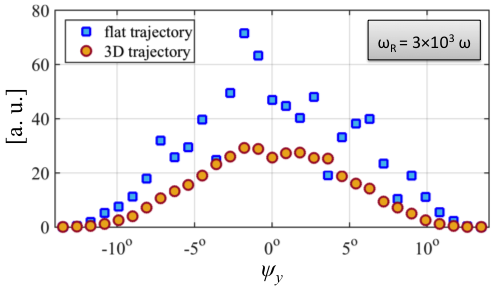}} 
  \caption{X-ray emission at frequency $\omega_R = 3 \times 10^3 \omega$. Upper panel: emission intensity as a function of the opening angles along the $x$ and $y$-axes ($\psi_x$ and $\psi_y$) for the trajectory shown in Fig.~ \ref{Figure2}. Lower panel: emission as a function of the opening angle in the plane of the driven oscillations $\psi_y$ (and summed over all $\psi_x$) for the 3D trajectory from Fig.~ \ref{Figure2}  and for a flat trajectory obtained for the same parameters, but no initial displacement.} \label{Figure_9}
\end{figure}

The development of the three-dimensional electron motion in the ion channel has strong implications for the x-ray emission by significantly increasing the divergence of the radiation. For  example, the upper panel in Fig.~\ref{Figure_9} shows the intensity of the x-ray emission as a function of the opening angles along the $x$ and $y$-axes ($\psi_x$ and $\psi_y$) for the three-dimensional trajectory from Fig.~ \ref{Figure2}. The pattern, shown for the x-rays with frequency $\omega_R = 3 \times 10^3 \omega$, is rather generic. It  remains similarly widened in the direction of the free oscillations ($x$-direction) for other frequencies and for the total spectral emission. In contrast with that, the emission pattern for an electron without any initial displacement along the $x$-axis would appear as a line with a width that scales as $1/\gamma$, where $\gamma$ is the characteristic $\gamma$-factor along the trajectory. However, along the $y$-axis (that is, in the plane of the driven oscillations), the width of the emission patterns for the three-dimensional and flat trajectories are comparable, as evident from the lower panel of Fig.~\ref{Figure_9} where the emission was summed over all $\psi_x$.

The three-dimensional electron motion also reduces the overall emitted energy. In the example shown in Fig.~\ref{Figure_9}, the energy is roughly halved when compared to the flat trajectory with the same axial length. We find that the local curvature of the electron trajectory tends to decrease due to the extra dimension. As a result, the acceleration decreases and the emission drops. The presented analysis indicates that the important physics might be missing in two-dimensional particle-in-cell simulations that are often used to predict the x-ray yield from plasma channels and hence three-dimensional simulations are likely necessary to correctly predict the x-ray emission by the electrons in this regime.

It has recently been shown that electron injection into the laser pulse can significantly lower the threshold for the electron energy enhancement~\cite{Arefiev2015}, while the superluminosity of the wave induced by the channel can have a detrimental effect of the electron energy gain~\cite{Robinson2015}. A comprehensive study is therefore necessary to determine the interplay of these aspects with the parametric instability. Nevertheless, the main effect of the spontaneous emergence of non-planar electron orbits during direct laser acceleration is triggered by strong modulations of the $\gamma$-factor and, because of that, it is not sensitive to the value of the phase velocity of the laser pulse. Finally, it should be pointed out that the direct laser acceleration is also being considered as a mechanism for boosting electron energies in the wakefield acceleration scheme~\cite{Zhang2015}. In principle, transverse electric fields generated by the bubble can have a destabilizing effect on the electron trajectories that is similar to the one we have discussed for the ion channels. It remains to  be seen whether the parametric instability in fact develops in the regimes favorable for the x-ray emission~\cite{Albert2013}. On the other hand, recent analytical and simulation results show that axially modulated plasma wakefields can be used to trigger the parametric instability in a controlled fashion, thus enhancing the betatron x-ray emission by the accelerated electrons~\cite{Palastro2015}. We have considered only axially uniform plasma channels. However, corrugated plasma channels with adjustable axial modulation periods can be reliably generated utilizing clusters and cluster plasmas~\cite{Layer2007, Layer2009}. This capability offers a possibility of controlling the electron motion utilizing the mechanism discussed in this paper. 


\section*{Acknowledgments}
This material is based upon work supported by the U.S. Department of Energy [National Nuclear Security Administration] under Award Number DE-NA0002723. A.V.A. was also supported by AFOSR Contract No. FA9550-14-1-0045, U.S. Department of Energy - National Nuclear Security Administration Cooperative Agreement No. DE-NA0002008, and U.S. Department of Energy Contract No. DE-FG02- 04ER54742. V.N.K. and G.S. were supported by AFOSR Contract No. FA9550-14-1-0045 and U.S. Department of Energy Contract Nos. DE-SC0007889 and DE-SC0010622.



\end{document}